\documentclass{optica-article}

\journal{opticajournal} 

\articletype{Research Article}
\usepackage{subcaption} 
\usepackage{lineno}
\usepackage{amsmath}
\usepackage{ulem}
\DeclareMathOperator{\Tr}{Tr}
\definecolor{Blue}{rgb}{0,0,0}
\newcommand{\DBR}[1]{{\color{Blue} #1}}
\definecolor{PurpleMagenta}{rgb}{0.75,0,0.75}
\newcommand{\LR}[1]{{\color{PurpleMagenta} #1}}
\definecolor{Red}{rgb}{0.8,0,0}
\newcommand{\NTM}[1]{{\color{Red} #1}}
\definecolor{Green}{rgb}{0,0.4,0}

\begin{document}

\title{Tunable passive squeezing of squeezed light through unbalanced double homodyne detection}

\author{Niels Tripier-Mondancin\authormark{1}, David Barral\authormark{1,2,*}, Ganaël Roeland\authormark{1}, Raúl Leonardo Rincon Celis\authormark{1}, Yann Bouchereau\authormark{1}, Nicolas Treps\authormark{1}}

\address{\authormark{1} Laboratoire Kastler Brossel, Sorbonne Universite, CNRS, ENS-Université PSL, College de France,
4 place Jussieu, F-75252 Paris, France\\
\authormark{2}Currently with the Galicia Supercomputing Center (CESGA), Avda. de Vigo S/N, Santiago de Compostela, 15705, Spain\\
}


\email{\authormark{*}david.barral.rana@gmail.com} 


\begin{abstract}
The full characterization of quantum states of light is a central task in quantum optics and information science. Double homodyne detection provides a powerful method for the direct measurement of the Husimi Q quasi-probability distribution, offering a complete state representation in a simple experimental setting and a limited time frame. \DBR{Here, we demonstrate that double homodyne detection can serve as more than a passive characterization tool.} By intentionally unbalancing the input beamsplitter that splits the quantum signal, we show that the detection scheme itself performs an effective squeezing or anti-squeezing transformation on the state being measured. The resulting measurement directly samples the Q function of the input state as if it were acted upon by a squeezing operator whose strength is a tunable experimental parameter: the beamsplitter's reflectivity. 
We experimentally realize this technique using a robust polarization-encoded double homodyne detection to characterize a squeezed vacuum state. Our results demonstrate the controlled deformation of the measured Q function's phase-space distribution, confirming that unbalanced double homodyne detection is a versatile tool for simultaneous quantum state manipulation and characterization.
\end{abstract}


    



\section{Introduction}

The ability to generate, manipulate, and precisely measure nonclassical states of light is a cornerstone of modern quantum technologies. Our capacity to probe the quantum world owes much to the development of homodyne detection, a technique pioneered in the 1980s that allows for the measurement of the quadratures of the electromagnetic field \cite{yuen1983, abbas1983}. These measurements are the bedrock of quantum state tomography, a set of methods for reconstructing quantum state representations, such as the Wigner function that is obtained from a series of quadrature measurements \cite{smithey1993, breitenbach1997}. 

While powerful, the reconstruction of Wigner functions of complex states from single-quadrature measurements can be notoriously demanding, requiring extensive data acquisition and sophisticated post-processing algorithms \cite{lvovsky2004, parigi2007}. To address these challenges, more direct characterization schemes have been pursued. An alternative is double homodyne detection (DHD), where a signal beam is split and simultaneously measured by two phase-locked homodyne detectors to probe orthogonal quadratures \cite{walker1984, walker1986}. The concept, which has deep roots in the quantum measurement theory of Arthurs and Kelly \cite{arthurs1965}, was shown to be a direct measurement of the Husimi Q function \cite{leonhardt1993, paris1996}. This approach bypasses tomographic reconstruction from Wigner marginal distributions and outperforms homodyne detection in moment-reconstruction accuracy for almost all Gaussian states and some important non-Gaussian states \cite{rehacek2015, muller2016,teo2017}. Moreover, the direct-access nature of this scheme enables significantly faster reconstruction of the Q function compared to standard Wigner-function tomography. This is especially useful in the multimode regime, because the required number of measurement settings does not grow exponentially with the number of modes \cite{roeland2023}. Remarkably, in a recent work microscopic Schrödinger cat tomographic data obtained by DHD has been used to emulate beyond-the-state-of-the-art mesoscopic Schrödinger cat breeding \cite{gottsch2025}.

Conventionally, the paradigm in quantum optics involves a clear separation of tasks: state generation --e.g., in a nonlinear crystal, subsequent manipulation --e.g., with beamsplitters or phase shifters, and finally, measurement. Each step is a distinct physical stage. The technique we explore here challenges this separation by integrating manipulation directly into the measurement apparatus itself. This concept of engineering the Positive Operator-Valued Measure (POVM) of a detector to perform a squeezing transformation is a powerful one, and its connection to unbalanced DHD has been previously identified theoretically \cite{chabaud2017}. By implementing this concept, we do not only simplify the optical setup but also introduce a powerful form of "on-line" state processing, where the measurement device is no longer a passive observer but an active participant that reshapes the quantum state it is designed to probe.

This capability for on-line squeezing is more than an experimental curiosity. It provides a continuously tunable knob to deform a state's phase-space distribution. Such a tool opens up intriguing possibilities for quantum metrology and quantum information science. For instance, we can dynamically reshape a state to better estimate a parameter encoded in the squeezing amplitude or phase of the input state \cite{pinel2013, tripier2025}. Similarly, we can tailor the state to maximize the fidelity to a given non-Gaussian state in a protocol for certification of high-order non-Gaussian features \cite{chabaud2021}. It effectively turns the detector into a simple, reconfigurable quantum processor, in which the squeezing operation is controlled on demand.

In this paper, we present the first, to our knowledge, experimental realization and application of this intrinsic squeezing transformation. We demonstrate, both in theory and practice, that by deliberately unbalancing the input beamsplitter --i.e., setting its reflectivity $R \ne 1/2$, the measurement apparatus itself becomes an active optical element acting as an effective squeezer, where the strength and orientation of the transformation are controlled by tuning the beamsplitter's reflectivity. The measurement outcome is no longer the Q function of the original state, but rather the Q function of that state after \DBR{undergoing} an effective squeezing transformation. To realize this concept, we rely on a robust polarization-encoded optical scheme which provides the necessary stability and tunability without a complex setup. We generate a multimode squeezed vacuum state from a synchronously pumped optical parametric oscillator and use our unbalanced DHD to directly observe the controlled deformation of its Q function. Our results show a clear reshaping of the input state in phase space, validating our theoretical model and establishing unbalanced DHD as a powerful and practical technique in the quantum optics toolkit.

\begin{figure}[htbp]
\centering\includegraphics[width=7cm]{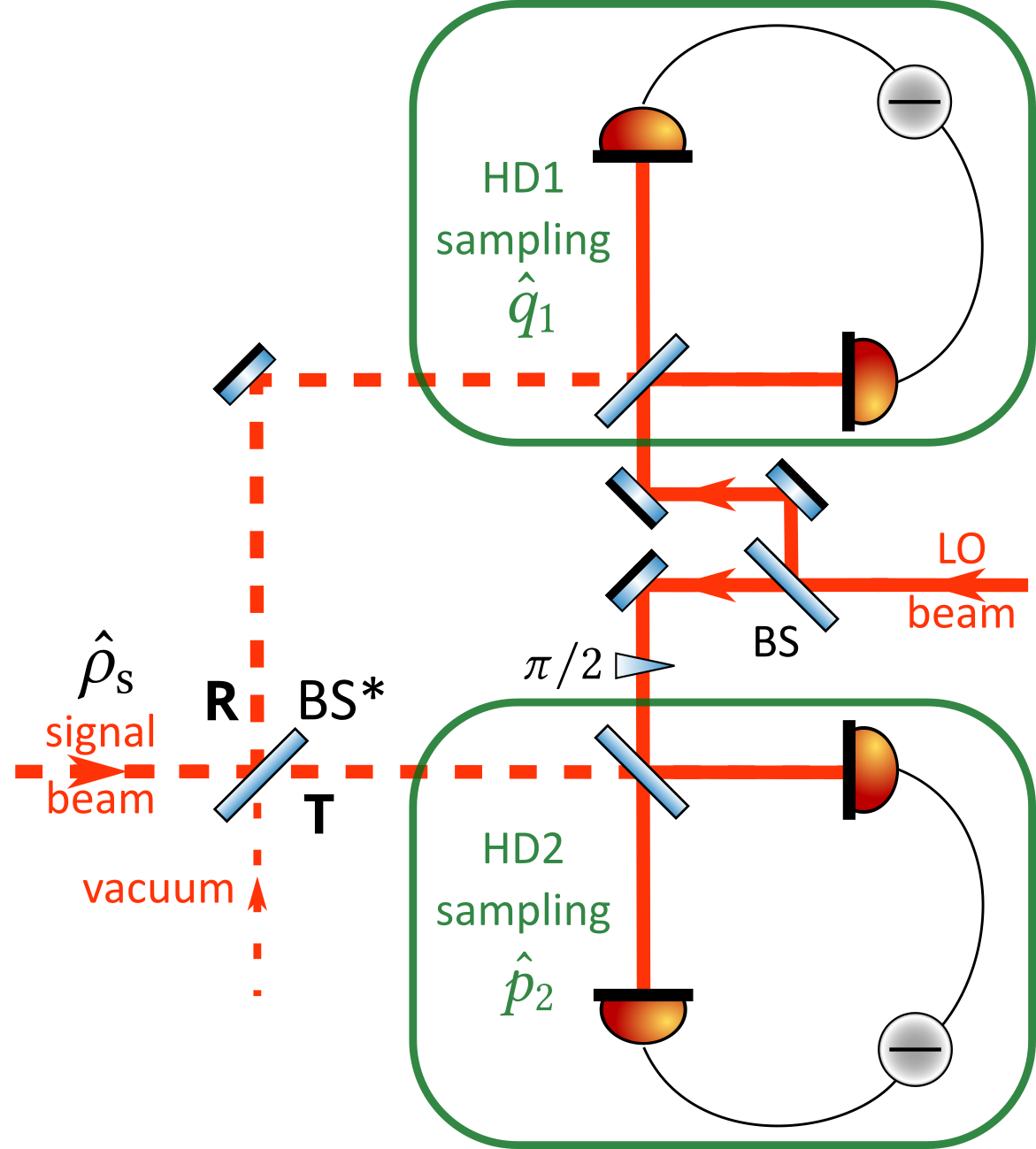}
\caption{Principle of operation of double homodyne detection. A signal state $\hat{\rho}_s$ is mixed with a vacuum field into the input beamsplitter BS${^*}$ with  reflectivity $R$. The two output fields are measured by two homodyne detectors HD1 and HD2 (green rectangles) with a phase reference given by two local oscillator (LO) with a relative phase of $\pi/2$. The two HDs thus sample orthogonal quadratures $\hat{q}_1$ and $\hat{p}_2$ of the input signal state with a precision limited by the Heisenberg uncertainty principle.}
\label{F1}
\end{figure}

\section{Double homodyne detection: theory}

A DHD system consists of two homodyne detection measuring different quadratures of the same state \cite{walker1986}. Figure \ref{F1} shows the principle of operation of the DHD: a single-mode quantum state $\hat{\rho}_s$ is split in two by the input beamsplitter, denoted BS$^{*}$, with reflectivity $R$. Then, homodyne detection is performed on each arm using two local oscillator dephased by $\pi/2$, ensuring the measurement of the two orthogonal quadratures of the light $\hat{q}(\theta)$ and $\hat{p}(\theta)$, where $\theta=\theta_{LO} - \theta_s$ is the relative phase between the signal beam and the LO, both in the same optical mode.

The precision in the simultaneous measurement of both quadratures is fundamentally limited by the Heisenberg uncertainty principle \cite{braunstein1991,stenholm1992}. This limitation is enforced by the mixing of the vacuum state with the signal beam at the input beamsplitter (BS$^*$), introducing vacuum noise into each path of the DHD. Consequently, each homodyne measurement includes vacuum fluctuations in the measured value.


In this work we focus on the Husimi Q function as DHD enables direct sampling of it. The Q function for a single-mode quantum state $\hat{\rho}$ is defined as \cite{leonhardt1997}
\begin{equation}
    Q_{\hat{\rho}}(\alpha)=\frac{1}{\pi} \langle \alpha \vert \hat{\rho} \vert \alpha \rangle,
\end{equation}
which is proportional to the projection of the quantum state on the coherent state $\vert \alpha \rangle$ with $\alpha \in \mathbb{C}$. Let us focus on the reconstruction of the Husimi Q function using DHD. The POVM of the \DBR{unbalanced} DHD is the set of projectors on squeezed displaced states given by \cite{roeland2023}
\begin{equation}
\hat{\Pi}_{DHD_{\xi}}(\alpha)=\frac{1}{\pi} \hat{S}(\xi)\hat{D}(\alpha)\vert 0 \rangle \langle 0 \vert \hat{D}(\alpha)^{\dagger} \hat{S}(\xi)^{\dagger},
\end{equation}
where $\hat{S}(\xi)=e^{\frac{\xi}{2}(\hat{a}^2 - \hat{a}^{\dagger 2})}$ is the single-mode squeezing operator and $\hat{D}(\alpha)=e^{\alpha \hat{a}^\dagger - \alpha^* \hat{a}}$ the displacement operator, with respective squeezing and displacement parameters given by
\begin{equation}
    \xi=\ln(\frac{r}{t})\in \mathbb{R},\qquad \alpha=\frac{1}{2}(\frac{q}{r}+i\frac{p}{t}) \in \mathbb{C},
\end{equation}
where $r,t$ are \DBR{respectively the reflection and transmission coefficients of the input beamsplitter BS$^{*}$, related to its reflectivity and transmitivity as $r^2 = R$, $t^2 = T$,} and where $r^2 + t^2 = 1$.

The probability of measuring the signal state $\hat{\rho}_s$ with outcome $\alpha$ is given by \cite{roeland2023}
\begin{equation}
P_{DHD_{\xi}}(\alpha)=\Tr(\hat{\rho}_s \hat{\Pi}_{DHD_{\xi}}(\alpha))=\frac{1}{\pi} \langle 0 \vert\hat{D}(\alpha)^{\dagger}  \hat{S}(\xi)^{\dagger} \hat{\rho}_s \hat{S}(\xi) \hat{D}(\alpha) \vert 0 \rangle= Q_{\hat{S}(-\xi)\hat{\rho}_s \hat{S}(-\xi)^{\dagger}}(\alpha). 
\end{equation} 
Thus, the probability distribution measured by \DBR{an unbalanced} DHD directly samples the Q-function of a squeezed signal state \DBR{$\hat{S}(-\xi) \hat{\rho}_s \hat{S}(-\xi)^{\dagger}$}. 

This squeezing operation is a unique feature of DHD, tuned by the input beamsplitter reflectivity $R$.  While balanced DHD ($R=1/2$) does not produce squeezing ($\xi=0$), unbalanced DHD  ($R \neq 1/2$) is formally equivalent to a squeezed operation followed by a balanced DHD. The squeezing factor in dB is given by
\begin{equation}\label{eq:unbalancingdB}
    s_{dB}=10\log_{10}(\frac{t^2}{r^2}).
\end{equation}
As $r$ tends to 0 or $1$, the squeezing parameter $\xi$ becomes infinite. In this limit the POVM of the double homodyne detection reduces to that of single homodyne detection, i.e. a set of projectors on infinitely squeezed states.

In practice,  building a histogram in the phase space using complex samples $\{ \alpha_i = \frac{1}{2}(\frac{q_{1}}{r}+i\frac{p_{2}}{t})\}$ from pairs of samples $\{q_{1},p_{2}\}$ measured at the two homodyne detectors, we are directly sampling the $Q$ function of a squeezed signal state.

In this paper, we study the effect of the unbalancing on a squeezed state. Let us take a single-mode pure squeezed state squeezed in the phase quadrature $\hat{\rho}_s=\hat{S}(s)\vert 0 \rangle\langle 0 \vert \hat{S}(s)^{\dagger}$ with squeezing parameter $s$. The Q-function of such a state is written as \cite{leonhardt1997}
\DBR{\begin{align}
    Q(q_1,p_2) &= \frac{1}{2\pi}\frac{1}{\sqrt{(1+s)(1+1/s)}}e^{-\frac{1}{(1+s)}\frac{q_1^2}{2}-\frac{1}{(1+1/s)}\frac{p_2^2}{2}},\label{eq:Qpure}
\end{align}}
where $q_1$ and $p_2$ are related to the displacement parameter as $\alpha = \frac{1}{\sqrt{2}}(q_1+ip_2)$. Through the unbalancing, the Q function of the measured state can now be written as 
\begin{align}
    Q'(q_1,p_2) &= Q(\frac{q_1}{\sqrt{2}r},\frac{p_2}{\sqrt{2}t}),\label{eq:Qnonpure}
\end{align}
where now $\alpha = \frac{1}{2}(\frac{q_1}{r}+i\frac{p_2}{t})$. For a non-pure state, the value of $s$ in each exponential of Equation \eqref{eq:Qpure} can be replaced by $s_s$ and $s_{as}$, the values of squeezing parameter and antisqueezing parameter.

\section{Double homodyne detection: polarization-based implementation}

From an experimental point of view, there are few challenges with the DHD scheme presented in Fig.\ref{F1}:
\begin{enumerate}    

    \item The LO-signal phase $\theta$ needs to remain constant over the full duration of the experiment to reconstruct the correct quantum state. Thus an active phase locking is essential. 
    \item The unbalancing carried out by the signal beamsplitter $BS^{*}$ needs to be adjustable easily. With current technology, using a usual beamsplitter is not possible because its coefficients of reflection and transmission at a given wavelength are fixed. It would require to change the optics every time one wants to apply a different unbalancing. 
    \item Implementing a fixed phase shift $\pi/2$ between the two HDs may be a challenging task as it would require to accurately lock the phase between the two LO beams of each HD to always measure orthogonal quadratures. This would add to the complexity of the locking system that fixes the LO-signal phase $\theta$, which may induce errors in the reconstruction of the Q function.
\end{enumerate}

\begin{figure}[htbp]
\centering\includegraphics[width=8cm]{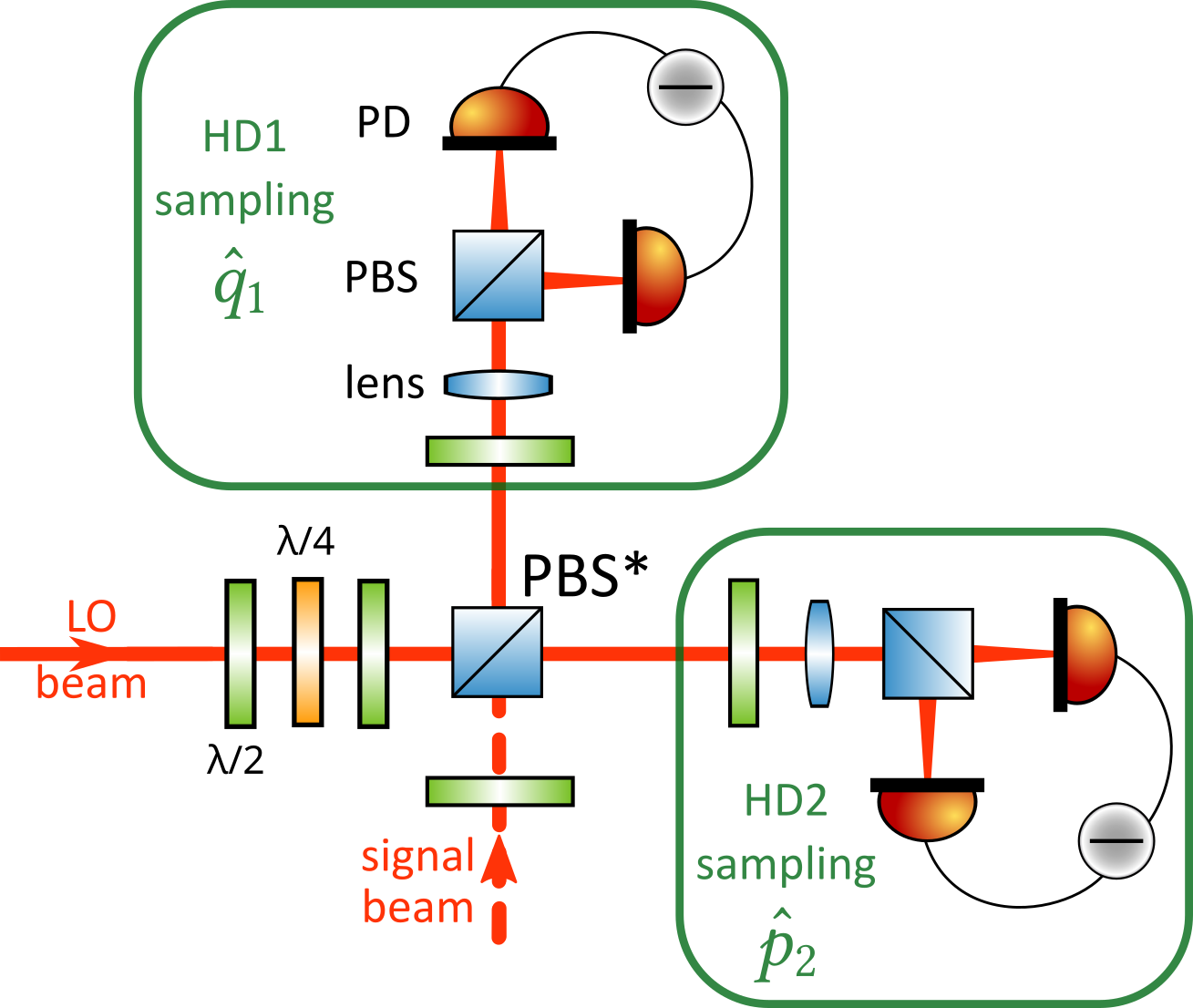}
\caption{Scheme of the experimental double homodyne detection setup based on polarization, where the input beamsplitter BS${^*}$ of Fig. \ref{F1} is substituted by a polarization beam splitter. A half-wave plate in the path of the signal beam is added to control the unbalancing and the local oscillator is prepared in circular polarization mode.}
\label{F2}
\end{figure}

The first issue is solved with an active LO-signal locking system commonly used in such experiments, through a hold and measure sequence. The details of this lock can be found in {\cite{roeland2023}}. 

The other two problems can be avoided relying on the polarization degree of freedom of the fields at play. The second issue is solved by replacing the input beamsplitter BS${^*}$ with a polarizing beamsplitter (PBS${^*}$) as shown in Fig. \ref{F2}. For a linearly-polarized input signal mode, a half-wave plate (HWP) followed by a PBS enables to tune the relative amount of power transmitted --horizontal polarization-- and reflected --vertical polarization-- by the PBS, and thus the unbalancing applied on the state. The third issue is solved using a circularly-polarized LO mode that naturally produces the $\pi/2$ phase shift between the two paths after the PBS${^*}$ (see {\cite{tripier2024}} for details). For a linearly-polarized LO mode this can be easily accomplished by a HWP followed by a quarter-wave plate (QWP). Adding another HWP element is also important if the total quantum efficiency of the two HDs is different, as a way to balance the quantum efficiency of both HDs. Moreover, taking advantage of the polarization encoding, we can also replace the respective BS at each HD by a system composed by a HWP and a PBS in order to fine tune and improve the balancing.

Taking the above into account, the measured quadratures at HD1 and HD2 are respectively
\DBR{\begin{align}
    \hat{q}_1(\theta, r) & \propto t \hat{q}_H (\theta) + r \hat{q}_V (\theta),\\
    \hat{p}_2(\theta, r) & \propto r \hat{p}_H (\theta) - t \hat{p}_V (\theta),
\end{align}}
where the subscripts H(V) stand for Horizontal(Vertical) signal polarization modes. Hence, the expected values of the quadratures measured by the DHD for a vertically-polarized input signal $\vert \psi\rangle \langle \psi \vert$ --with the horizontally-polarized mode in vacuum $\vert 0 \rangle \langle 0 \vert$-- are 
\DBR{\begin{align}
    \langle\hat{q}_1(\theta, r) \rangle &= r  \langle\hat{q}_V (\theta)\rangle,\\
   \langle \hat{p}_2(\theta, r) \rangle &=- t \langle \hat{p}_V (\theta)\rangle. \label{is}
\end{align}}
As expected, the measured quadrature from HD1 is proportional to the amplitude quadrature whereas that of HD2 to the phase quadrature. The scaling factors depend on the unbalancing $\{r,t\}$. The respective variances are
\DBR{\begin{align}
    \langle\Delta^{2} \hat{q}_1(\theta, r)\rangle = r^2 \langle\Delta^{2} \hat{q}_V(\theta)\rangle + t^2, \\
    \langle\Delta^{2} \hat{p}_2(\theta, r)\rangle = t^2 \langle\Delta^{2} \hat{p}_V(\theta)\rangle + r^2.
\end{align}}
The variance of the quadrature measured in HD1 is now proportional to a squeezed amplitude-quadrature variance and that from HD2 to a antisqueezed phase-quadrature. The effect of input vacuum fluctuations in the DHD becomes evident here as an added factor.

\section{Double homodyne detection: experiment}

In this section, we present the details of our experimental setup composed of a source of squeezed states and a DHD acquisition system.

\subsection{Source of squeezed states}

\begin{figure}[htbp]
\centering\includegraphics[width=12cm]{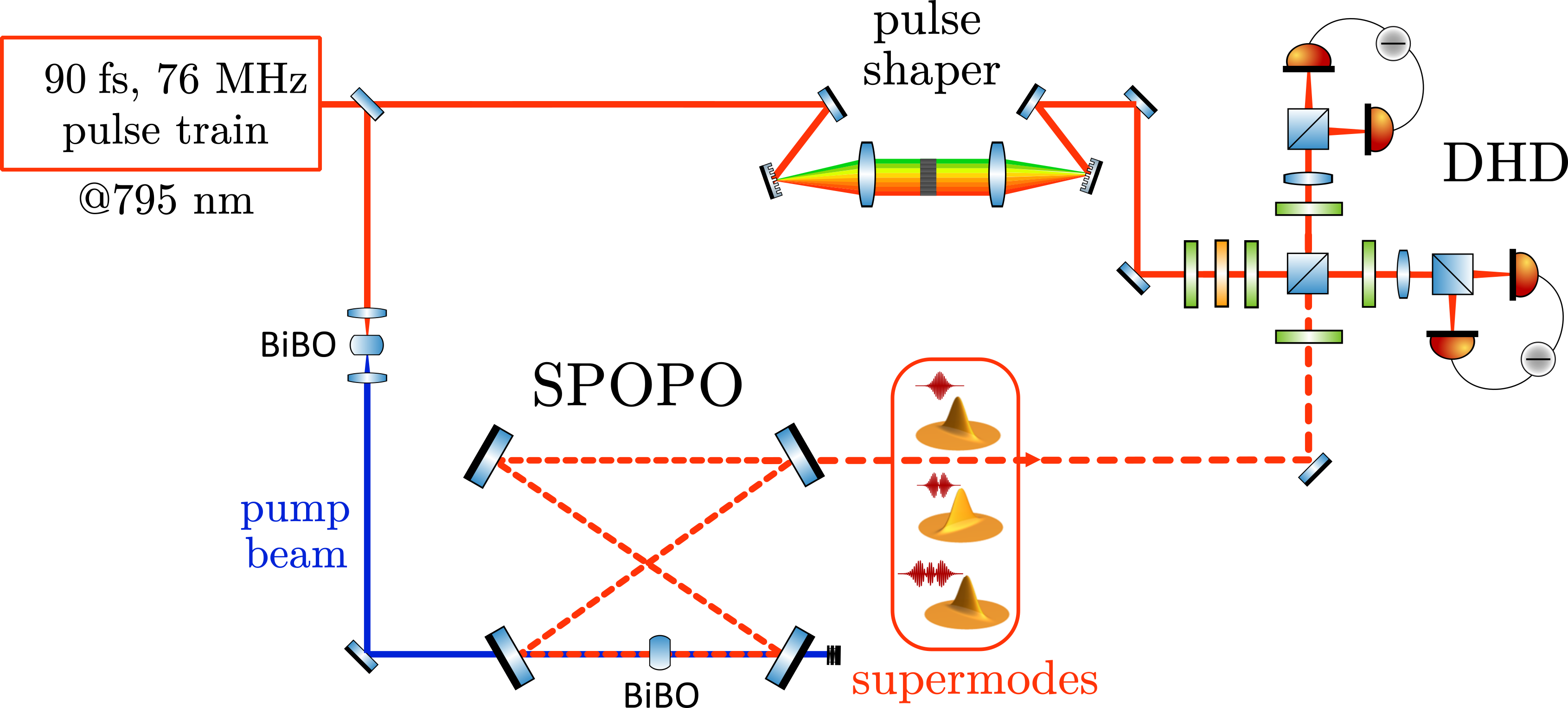}
\caption{Scheme of the experimental setup. The initial beam is split in two parts. The first part is used for the generation of the state. A SPOPO cavity is pumped by an up-converted pump beam, resulting in a multimode squeezed vacuum at the output. The second part is used for the local oscillator of the DHD, where we can chose the mode of the LO using a pulse shaper.}
\label{F5}
\end{figure}

The light source of our experimental setup is a Ti:Sapphire laser (Coherent Mira 9000-F) which produces a train of approximately Gaussian pulses centered at 795 nm with a duration of 90 fs at repetition rate of 76 MHz (Fig. \ref{F5}). This beam is split in four: one pumps an optical parametric oscillator (OPO) after up-conversion, two are used for phase-lock and alignment purposes --cavity lock and seed beams-- and one as a tunable LO  that sets the measurement basis of the DHD scheme. The first beam is up-converted to a center wavelength of 397.5 nm with a full width half maximum (FWHM) of 3 nm that pumps a 2-mm nonlinear BiBO crystal (Castech) placed in a low-finesse synchronously pumped OPO (SPOPO) cavity \cite{roslund2014}. The pump power is set at 50 mW and the output-coupler transmittance at 50\% to ensure the high purity of the output state --at the cost of a reduced \DBR{parametric downconversion} gain. The cavity outputs a multimode squeezed vacuum with approximately 10 Hermite-Gaussian spectral eigenmodes and a FWHM of $\approx 10.3$ nm in the first eigenmode. 

\subsection{Data acquisition}

We apply a hold-and-measure scheme blocking the seed by a mechanical chopper (Thorlabs MC2000B) at a 3 kHz rate where we alternatively lock the LO-signal phase $\theta$ and then measure homodyne traces from the signal beam. We measure for 118 $\mu s$, counting for 35\% of the duty cycle. The DHD setup based on polarization sketched in the previous section is mounted using low-loss PBSs (Edmund Optics) and zero-order waveplates (CVI laser optics). 

In each experiment we acquire 10 million samples from each homodyne $\{\hat{q}_1(\theta, r),\hat{p}_2(\theta, r)\}$ over 5 seconds using an oscilloscope (Teledyne Lecroy Waverunner 6Zi). As our squeezing spreads among several tens of individual pulses in the cavity, we need to apply the temporal mode of the cavity $\psi(t)$ to the raw data to obtain quadrature values related to that squeezed mode. We retrieve this mode from the temporal autocorrelation function of the homodyne signal and a principal component analysis (PCA) \cite{nielsen2006, morin2013}. $\psi(t)$ is approximated by a double decaying exponential function with a full width half maximum of 80 ns. We use this mode to reconstruct the quadratures associated to the squeezed state from the quadratures sampled by the detectors $\hat{q}_{1,2}(\theta, r, t_j)$ as \cite{ra2020}
\begin{align*}
    \hat{q}_{1,2}(\theta, r)= \sum_{j=1}^N  \psi(t_j) \hat{q}_{1,2}(\theta, r, t_j) \Delta t,  
\end{align*}
where $\Delta t$ corresponds to the inverse of the sampling rate, and $\hat{q}_{1,2}(\theta, r)$ is the quadrature value that we use to reconstruct the squeezed Q function. Each quadrature value comes from \DBR{segments} of 2 $\mu$s homodyne signal taken at a sampling rate of 100 MHz ($\Delta t=10$ ns) resulting in $N=200$ \DBR{windows} where we apply a discretized temporal mode $\psi(t_j)$ of equal length. The total number of pairs $\{\hat{q}_1(\theta, \phi),\hat{p}_2(\theta, \phi)\}$ used for a single Q-function reconstruction is $N_p=5.10^4$. \DBR{To perform the histogram, each axis of the phase space is divided into 100 bins between -3 and +3 $\sqrt{{\small SNU}}$, where SNU stands for shot noise units}.

Our home-made HDs present a common mode rejection ratio (CMRR) of -32 dB and 16 dB of electronic-to-shot-noise clearance over a 40 MHz bandwidth, featuring a detection quantum efficiency of $\eta_{e}=0.94$. The efficiency of the HD photodiodes (Hamamatsu S3883-02) at our working wavelength is $\eta_{p}=0.96$. The spatial overlap between signal and LO is measured as $96\%$, giving an effective efficiency of $\eta_{s}=0.92$. The total optical transmission related to waveplates, lenses, mirrors, etc. is $\eta_{o}=0.96$. The total efficiency of the homodyne detection is then given by the product of the above efficiencies as $\eta_{HD}=\eta_{e} \,\eta_{p}\, \eta_{s}\, \eta_{o}=0.80$. 

Our DHD setup enables the full characterization of the multimode squeezed vacuum generated at the SPOPO. However, in this work we are just interested in a single mode. We measure the first supermode of the cavity HG$_0$ by shaping the LO in that spectral mode. This is accomplished by means of a pulse shaper arranged in a folded $4-f$ line composed of a grating, a cylindrical mirror and a spatial light modulator (SLM, Hamamatsu LCOS-SLM X15213-02) \cite{cai2017}.

\section{Results}

\begin{figure}[h!]
	\centering\includegraphics[width=0.85\textwidth]{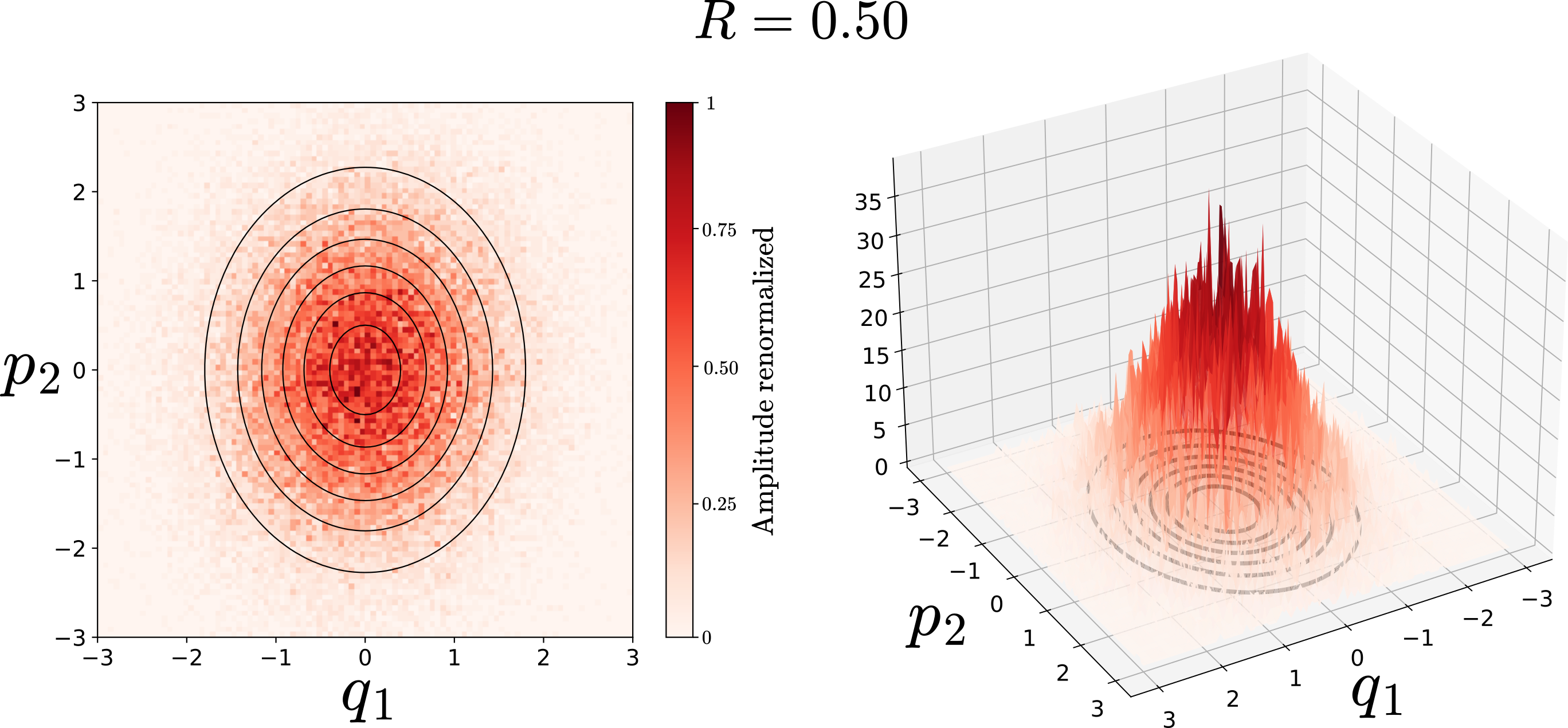}
	\caption{Experimental Q function of a squeezed state reconstructed with $N_p=5\times 10^4$ samples. The black lines represent the contours of the fit by a two-dimensional Gaussian. The fitted values give a squeezing and antisqueezing levels of -1.25 dB and 2.6 dB, respectively.}
	\label{F3}
\end{figure}

We first reconstruct the Q function of the input squeezed state using a balanced DHD with $R=1/2$ (Fig. \ref{F3}). By fitting the data with a two-dimensional Gaussian, we can recover the values of squeezing and antisqueezing of the original quantum state. Here, the fit gives -1.25 dB and 2.6 dB as squeezing and antisqueezing levels. These are raw values without correcting channel and detection losses that degrade the purity of the measured quantum state. \DBR{Remarkably, our Q-function reconstruction --quantum state tomography-- is done in a few seconds.} 

\begin{figure}[h!]
 \centering
\begin{tabular}{cc}
     \includegraphics[width=0.44\textwidth]{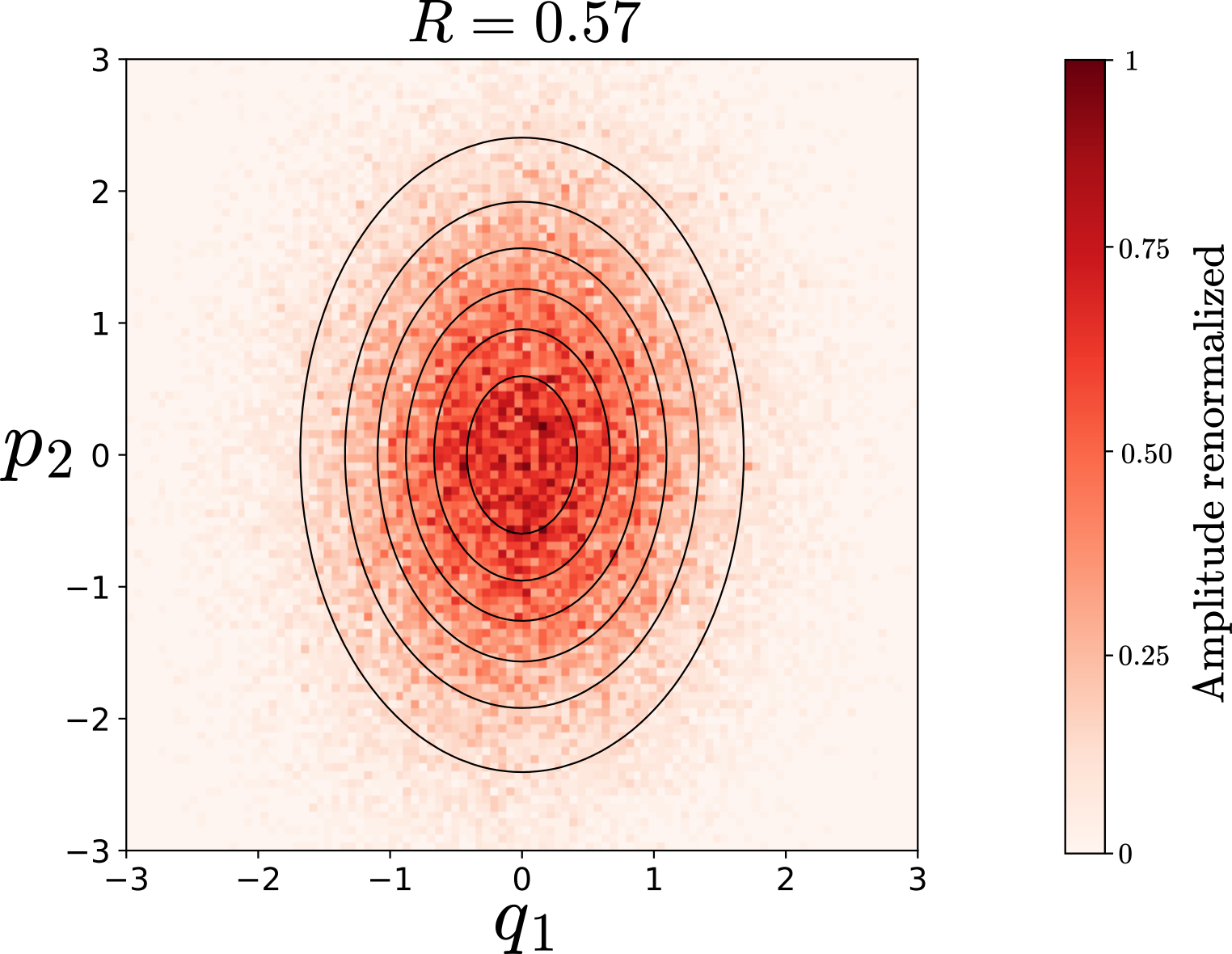} & 
  \includegraphics[width=0.32\textwidth]{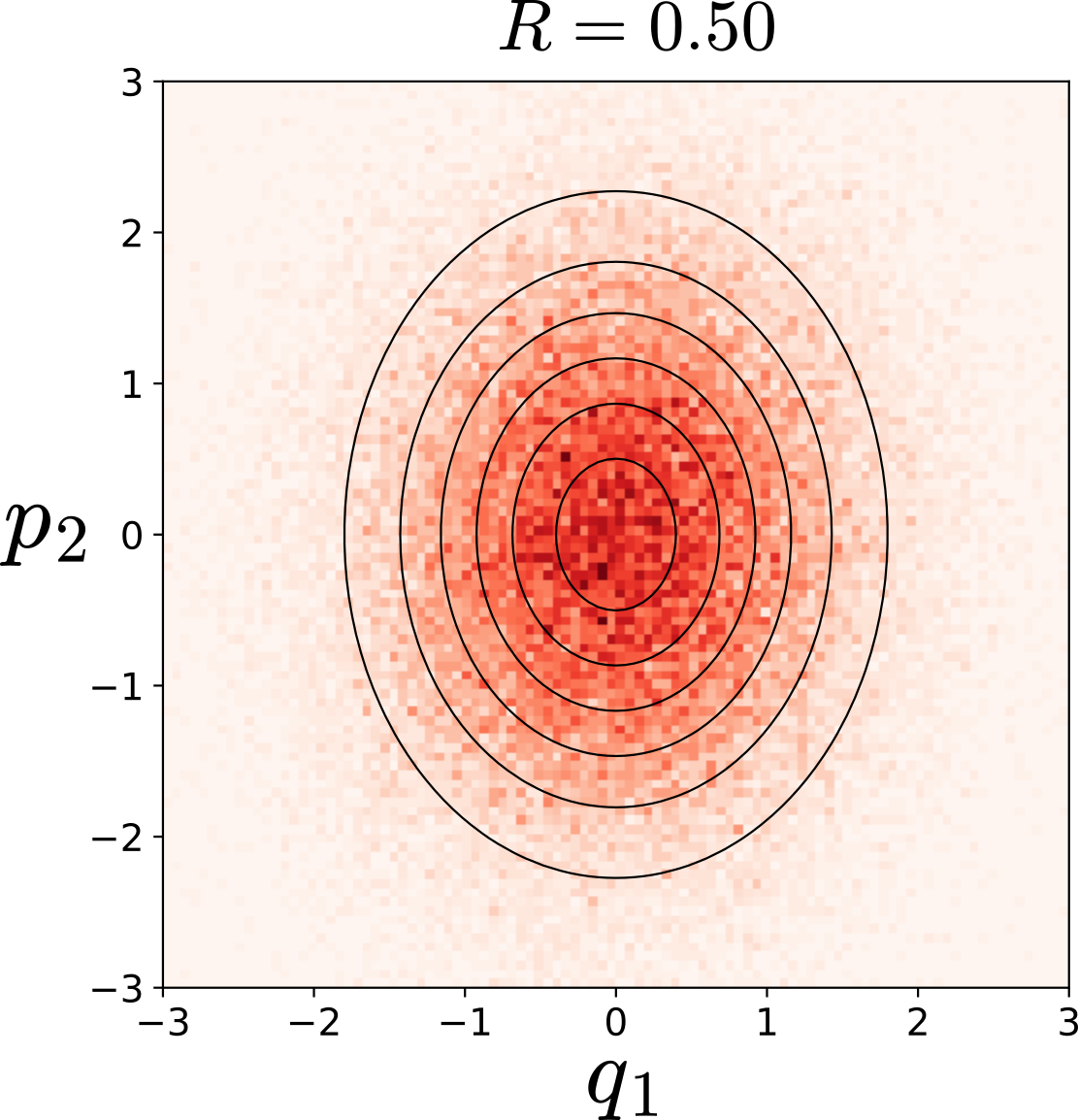}  \vspace{.4cm}\\
\includegraphics[width=0.44\textwidth]{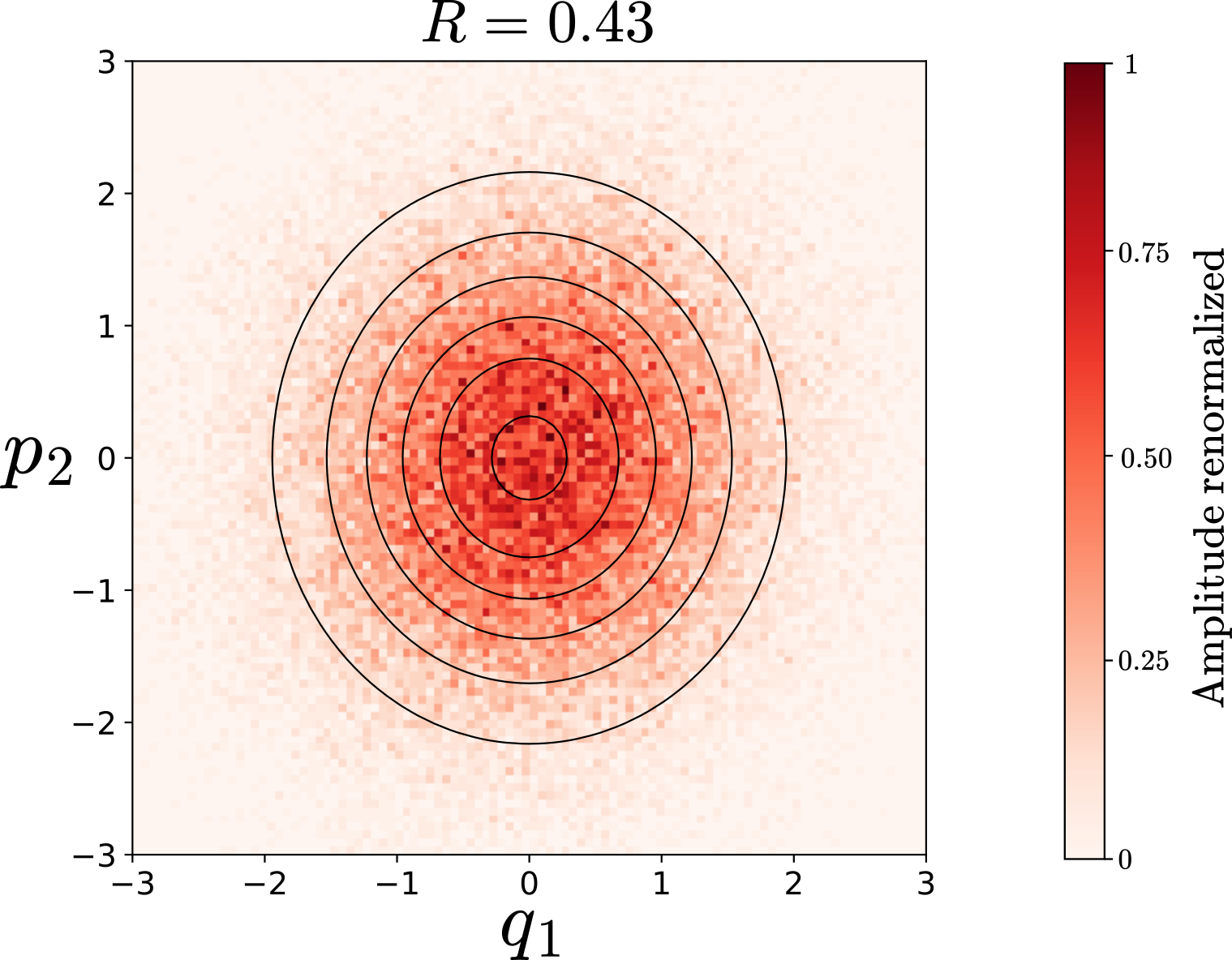} & 
  \includegraphics[width=0.32\textwidth]{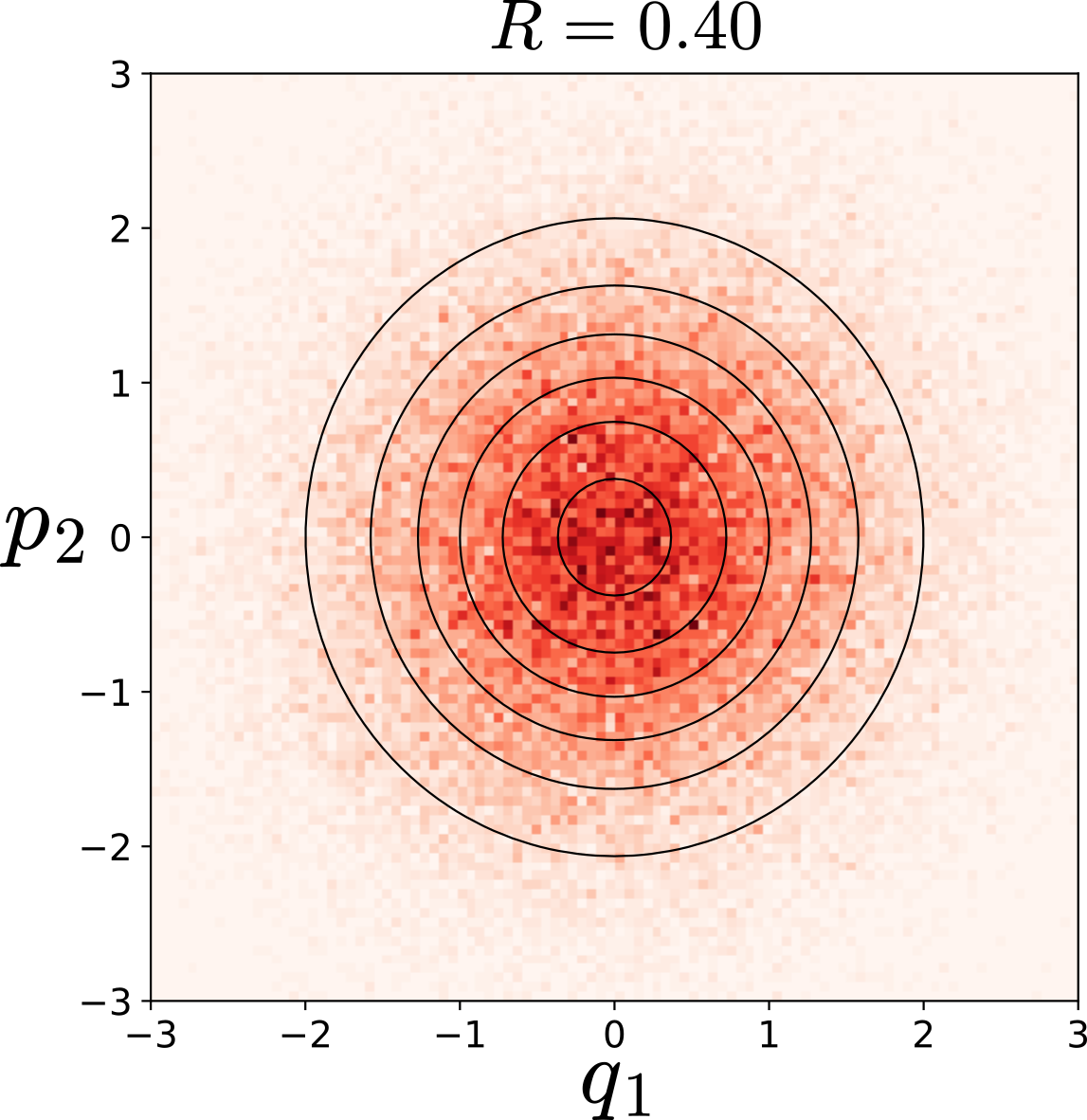}  \vspace{.4cm}\\
 \includegraphics[width=0.44\textwidth]{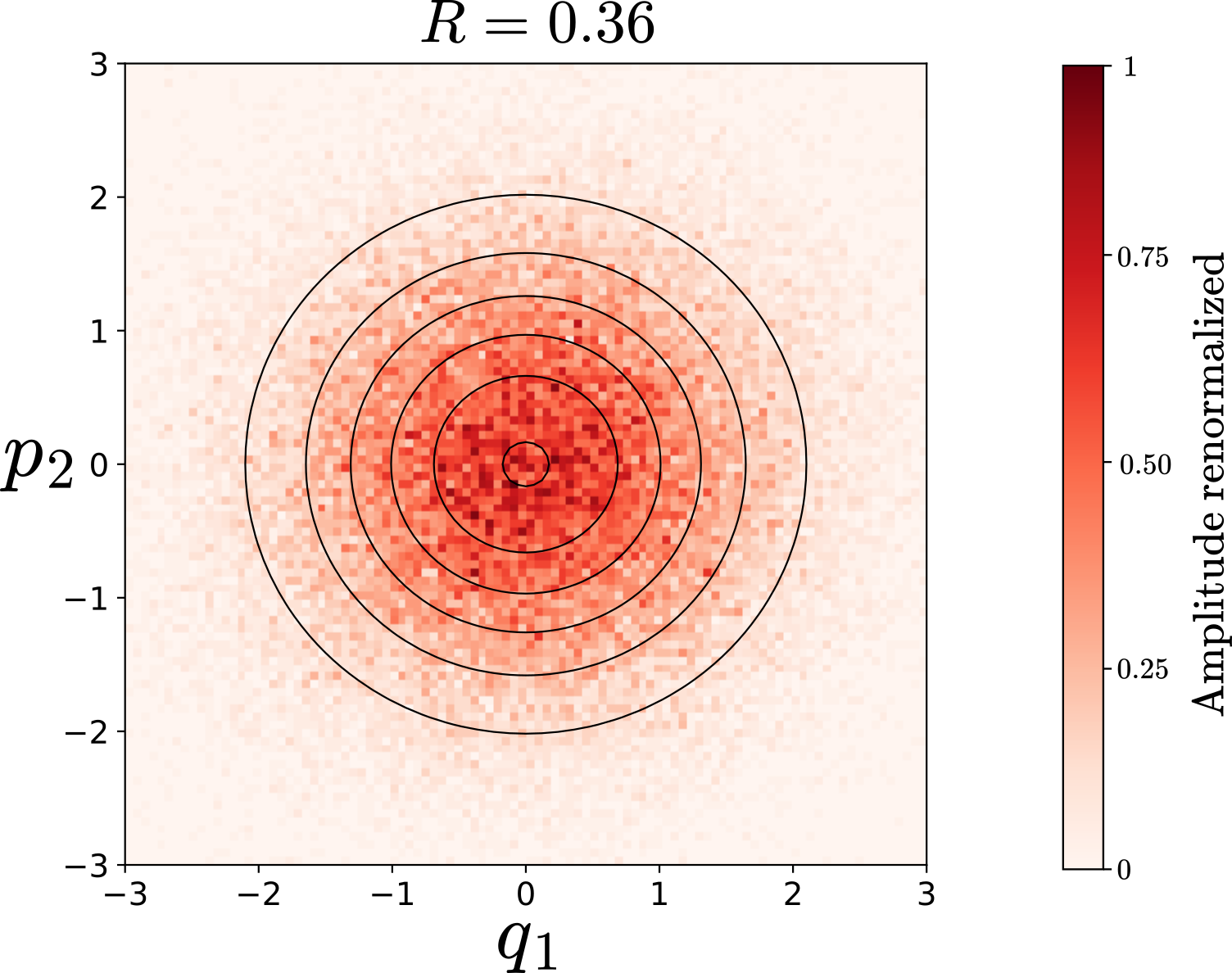} & 
  \includegraphics[width=0.32\textwidth]{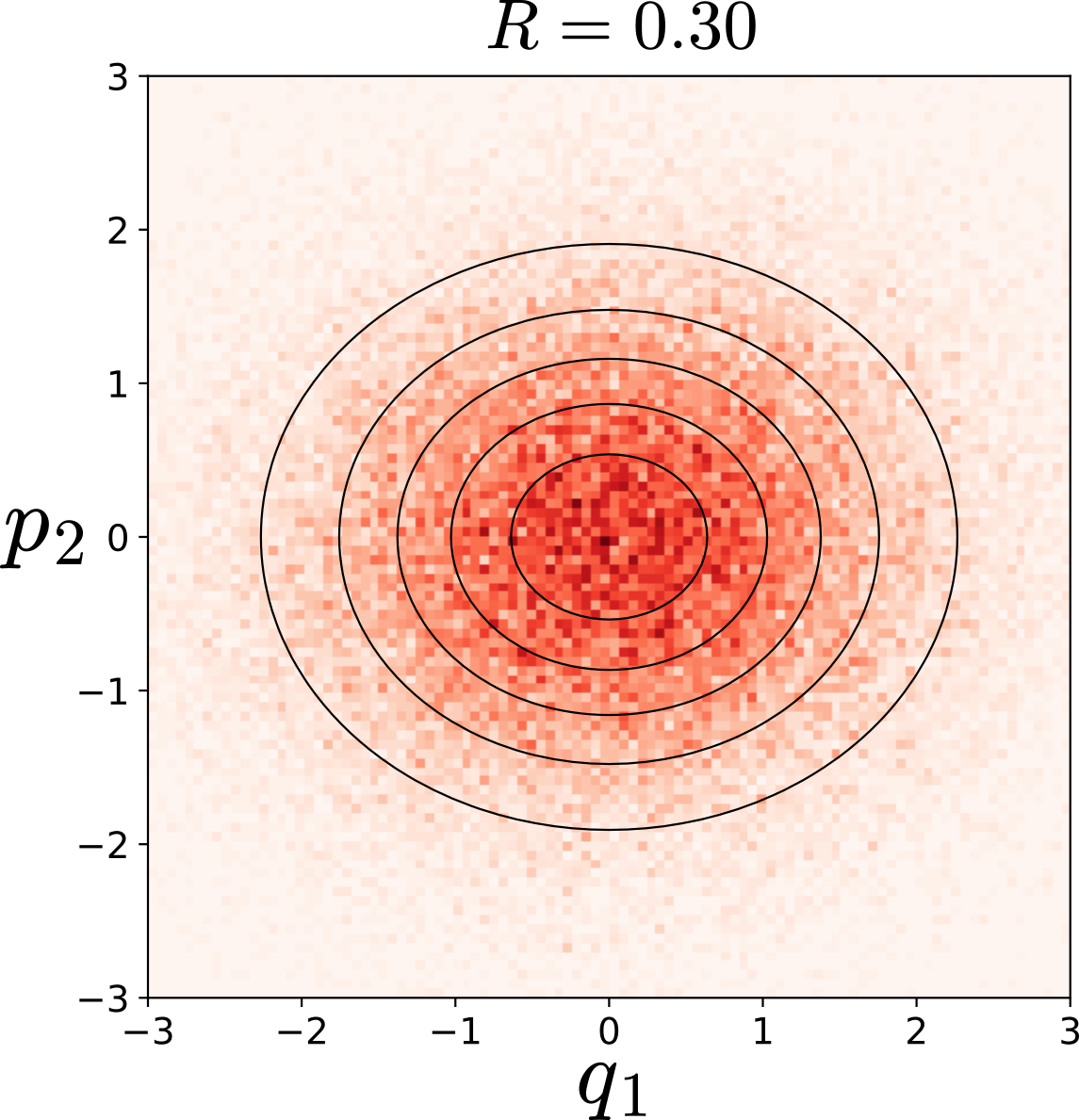}  \vspace{.4cm}\\
 \end{tabular}\\
    \caption{Experimental Q function of a squeezed state for different values of R reconstructed with $N_p=5\times 10^4$ samples. Values of R above 0.5 squeeze the state in the amplitude quadrature $q_1$ whereas values of R below 0.5 along phase quadrature $p_2$ in turn.}
\label{fig:Qunbalancing}
 \end{figure}

Next, we unbalanced the DHD varying $R$. Fig. \ref{fig:Qunbalancing} shows the reconstructed Q functions for six different reflection coefficients. The reconstructed Q functions exhibit the effect of unbalancing: by going to reflectivity R above $0.5$, we keep squeezing the state in the $q$ quadrature, whereas decreasing the reflectivity below 0.5 we squeeze the state in the $p$ quadrature. Notably, for $R = 0.4$ we fully unsqueeze the state --both quadrature variances are equal-- obtaining a Q function that resembles that of a thermal state.

\begin{figure}[htbp]
\centering\includegraphics[width=8cm]{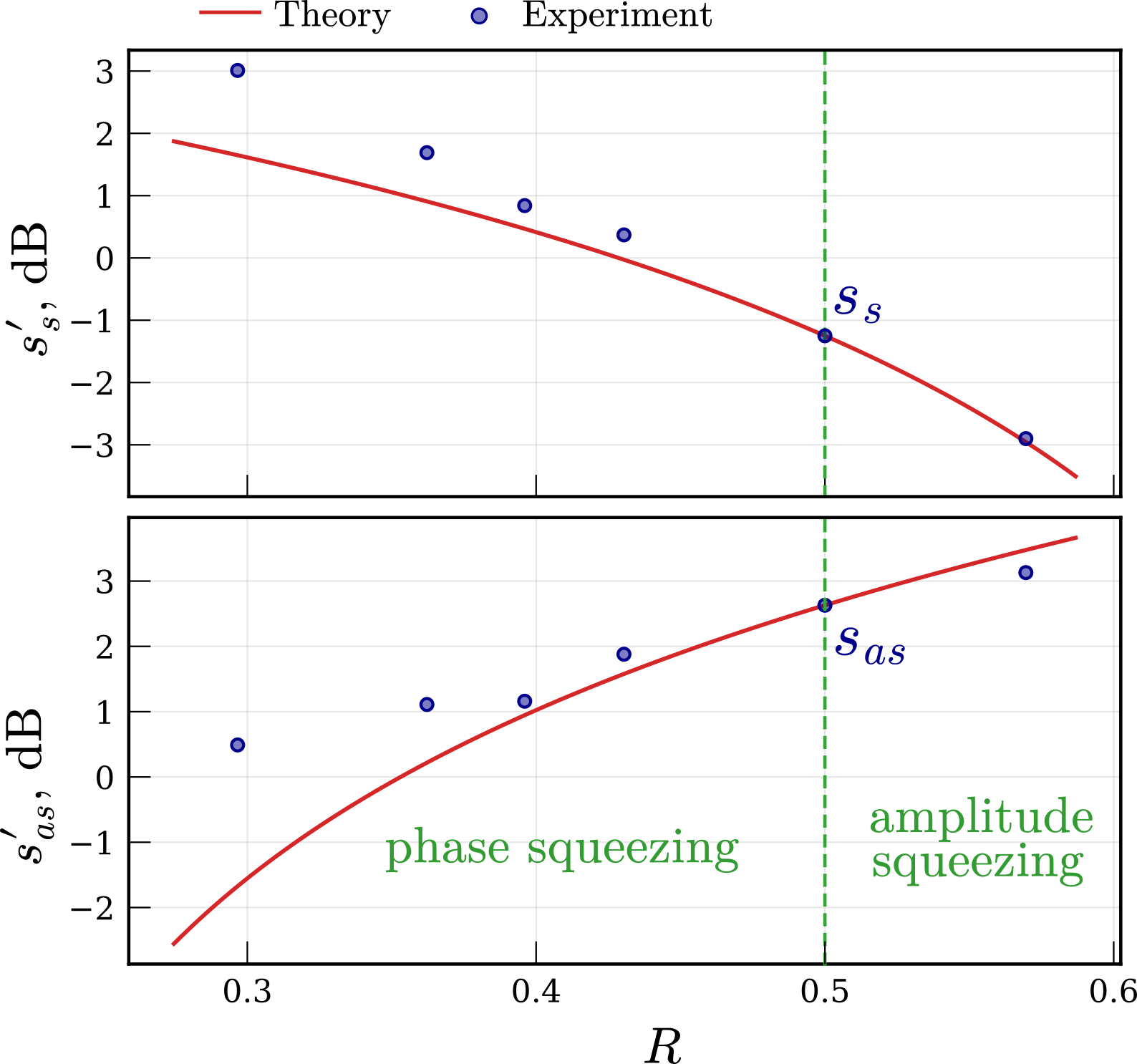}
\caption{Squeezing $s'_{s}$ and antisqueezing $s'_{as}$ of the fitted Q function upon unbalancing of the squeezed state. The squeezing of the initial state is found at $R = 0.5$. The two regimes using unbalancing are highlighted in green.}
\label{fig:sqz_asqz}
\end{figure}

The quantum states generated in our SPOPO cavity are not pure squeezed states, but thermal squeezed states. Using Equation \eqref{eq:Qnonpure} for a single-mode thermal squeezed state, the parameters of squeezing and antisqueezing of the fitted Q function ($s'_{s}$, $s'_{as}$) can be related to the parameters of the unbalancing ($r$,$t$) and the original squeezing of the state ($s_{s}$, $s_{as}$) as
\begin{align}
    s'_{s} &= 2r^2(1+s_s)-1, \\
    s'_{as} & = \Bigl(2t^2(1+\frac{1}{s_{as}})-1\Bigr)^{-1}.
\end{align}
Figure \ref{fig:sqz_asqz} shows the comparison between the experiment and the theory for the six measured experimental points. The first  point ($R=0.3$) diverges a bit from the theory, which can be explained by the fact that this point was measured last, and the stability of the whole experiment did not allow for a perfectly similar squeezed state to be produced --while each single experiment took 5 seconds, data transfer from the oscilloscope and preparing the next measurement extended the total time to $\approx 2$ minutes per Q function.
The green dashed line highlights the two different regimes in which we operate: phase squeezing and amplitude squeezing. For $R=0.5$ we reconstruct the original state, measuring its squeezing. One specific value of $R$ stands out in this experiment: the one for which the state is completely unbalanced. The unbalancing applied in dB is given by Equation \eqref{eq:unbalancingdB}, and needs to be equal to the  semidifference of the measured values in dB of the squeezing and antisqueezing. In this case, it corresponds to $s_{dB} = -1.925$ dB and $R=0.39$. 
This is consistent with the experiment, as is visually shown in Figure \ref{fig:Qunbalancing} for the corresponding value or $R$.


\section{Conclusions}

\DBR{In this work, we have demonstrated that double homodyne detection can operate as more than a passive characterization tool. By implementing a tunable, polarization-encoded experimental setup, we successfully performed an effective squeezing transformation on an input quantum state directly within the measurement apparatus.}

Our results confirm that intentionally unbalancing the input beamsplitter in a double homodyne detection scheme allows for the controlled deformation of the measured Husimi Q function. We experimentally verified this by measuring a squeezed vacuum state generated by a synchronously pumped optical parametric oscillator, demonstrating the ability to continuously tune the detected state from phase-squeezed to amplitude-squeezed regimes. Notably, we showed the capability to {\it unsqueeze} the input state, effectively recovering a thermal state-like Q function at a specific unbalancing ratio. The strong agreement between our experimental data and the theoretical model validates the versatility of this technique. 

\DBR{Double homodyne detection provides a robust, resource-efficient and tomographically-complete alternative to complex quantum-state reconstruction techniques \cite{lvovsky2009}, that enhanced with effective squeezing unbalancing may further improve advanced characterization protocols such as certification of non-Gaussianity and Wigner negativity in photon-subtracted squeezed states \cite{chabaud2021, tripier2024b}.}

 \newpage

\begin{backmatter}
\bmsection{Funding}
This work was supported by QuantERA II project SPARQL that has received funding from the European Union’s Horizon 2020 research and innovation programme under Grant Agreement No 101017733. \DBR{We acknowledge fundings from the HORIZON-EIC-2022-PATHFINDERCHALLENGES-01 programme under Grant Agreement Number 101114899 (Veriqub). Views and opinions expressed are however those of the authors only and do not necessarily reflect those of the European Union. Neither the European Union nor the granting authority can be held responsible for them.} D.B. also acknowledges the support by MICINN through the European Union NextGenerationEU recovery plan (PRTR-C17.I1), the Galician Regional Government through “Planes Complementarios de I+D+I con las Comunidades Autónomas” in Quantum Communications.


\bmsection{Disclosures}
The authors declare no conflicts of interest.

\bmsection{Data Availability Statement}
Data underlying the results presented in this paper are not publicly available at this time but may be obtained from the authors upon reasonable request.
\end{backmatter}


\end{document}